\documentclass[10pt,aps,prd,onecolumn,showpacs,amsmath,amssymb,nofootinbib,eqsecnum,preprintnumbers,superscriptaddress]{revtex4-2}

\usepackage{mathtools}					
\usepackage{xcolor}

\usepackage[colorlinks=true,
            citecolor=red,
            linkcolor=blue,
            urlcolor=violet,
            filecolor=cyan,
            backref=false]{hyperref}

\begin{document}


\title{Solutions with pure radiation and gyratons in 3D massive gravity theories}

\author{Ercan Kilicarslan}
\email{ercan.kilicarslan@usak.edu.tr}
\affiliation{Department of Mathematics, Usak University, 64200, Usak, Turkey}

\author{Ivan Kol\'a\v{r}}
\email{ivan.kolar@matfyz.cuni.cz}
\affiliation{Institute of Theoretical Physics, Faculty of Mathematics and Physics, Charles University,
V Hole\v{s}ovi\v{c}k\'ach 2, Prague 180 00, Czech Republic}

\date{\today}

\begin{abstract}
We find exact solutions of topologically massive gravity (TMG) and new massive gravity (NMG) in ${2+1}$ dimensions (3D) with an arbitrary cosmological constant, pure radiation, and gyratons, i.e., with possibly non-zero $T_{uu}$ and $T_{ux}$ in canonical coordinates. Since any `reasonable' geometry in 3D (i.e., admitting a null geodesic congruence) is either expanding Robinson-Trautman ($\Theta\neq0$) or Kundt (${\Theta=0}$), we focus on these two classes. Assuming expansions ${\Theta=1/r}$ (`GR-like' Robinson-Trautman) or ${\Theta=0}$ (general Kundt), we systematically integrate the field equations of TMG and NMG and identify new classes of exact solutions. The case of NMG contains an additional assumption of $g_{ux}$ being quadratic in $r$, which is automatically enforced in TMG as well as in 3D GR. In each case, we reduce the field equations as much as possible and identify new classes of solutions. We also discuss various special subclasses and study some explicit solutions.
\end{abstract}

\maketitle


\section{Introduction}

It is well known that in \textit{${2+1}$ dimensions (3D)} the \textit{general relativity (GR)} does not have any local propagating degrees of freedom. Nevertheless, despite suffering from the local triviality, the 3D GR augmented with a cosmological constant has a globally non-trivial structure with a healthy holographic conformal field theory that has two copies of the Virasoro algebra having positive central charges. Unlike the 3D GR, the \textit{topologically massive gravity (TMG)} \cite{Deser:1982vy} and the \textit{new massive gravity (NMG)} \cite{Bergshoeff:2009hq} have local dynamical propagating degrees of freedom. TMG is both unitary and renormalizable \cite{Deser:1990bj} and parity non-invariant (with a single helicity mode) theory with a single massive spin-2 particle about its flat or (anti)-de sitter [(A)dS] spacetimes. NMG is a unitary and parity-preserving theory with massive spin-$2$ excitations with helicities $\pm2$. These theories have bulk-boundary unitarity clash which has been a difficulty in giving a consistent unitary dual two-dimensional conformal field theory for bulk gravity in anti-de Sitter spacetime. It was also shown in \cite{Alkac:2017vgg} that $f(R)$ extension of 3D massive gravity theories does not cure the situation. On the other hand, causality and unitarity are compatible in these theories \cite{Edelstein:2016nml} once the sign of the Einstein-Hilbert term in the action is chosen negative as opposed to the 4D case, unlike the Einstein-Gauss-Bonnet or cubic theories in higher dimensions \cite{Camanho:2014apa}.

Exact solutions of GR corresponding to spinning null beams were first discovered by Bonnor in 1970. At the linearized level, the solutions for spinning ultrarelativistic particles were later studied by Frolov and Fursaev \cite{Frolov:2005in}, and called \textit{gyratons}. This term was later extended to various other exact solutions describing spinning null matter propagating in Minkowski background within the pp-wave class \cite{Frolov:2005zq,Frolov:2005ja,Podolsky:2014lpa} (see also \cite{Griffiths:2009dfa}) or in (anti-)de Sitter background \cite{Frolov:2005ww}, Melvin universe \cite{Kadlecova:2010je}, or direct product spacetimes \cite{Kadlecova:2009qu,Krtous:2012qa} within a more general Kundt spacetimes \cite{Frolov:2005ww,Podolsky:2008ec}, which has vanishing expansion of a null geodesic congruence. The gyratons were also discovered within the Robinson-Trautman spacetimes \cite{Podolsky:2018oov, Svarc:2014noa}, which has the non-zero expansion. Most of the above results were derived in arbitrary dimensions including the 3D GR. It should be stressed that the word `gyraton' is abused in the context of 3D spacetime. It only means that apart from the pure radiation $T_{uu}\neq0$ (in canonical coordinates), we also admit $T_{ux}\neq0$ (in analogy to $T_{ux}\neq0$ and $T_{uy}\neq0$ in four dimensions). Nevertheless, since the transverse space is 1-dimensional in 3D (and spanned by the coordinate $x$), $T_{ux}\neq0$ cannot represent any rotating matter. The actual interpretation of gyratons in 3D is unknown. \footnote{In 4D, the components $T_{ux}$ and $T_{uy}$ are chosen so that they combine into an angular momentum in the transverse space. In 3D, however, $T_{ux}$ may only represent a linear momenta in the transverse space, so it is possible that these solutions might be interpreted as tachyons.} Beyond GR, gyratons were also studied in higher-derivative (as well as infinite-derivative) theories of gravity in four dimensions \cite{Kolar:2021uiu}. 

Many works have been devoted to understanding the physical properties of 3D massive gravity theories in the context of exact solutions, see e.g. \cite{garcía-díaz_2017} (and references therein). Hence, various exact solutions of TMG and NMG were identified. Algebraic classification of exact solutions of cosmological TMG was studied systematically in \cite{Chow:2009km} where it was shown that most of the solutions known in the literature are locally equivalent to timelike-squashed AdS$_3$ or spacelike-squashed AdS$_3$ or AdS pp waves. Except for these three solutions, in the absence of cosmological constant, triaxially squashed AdS$_3$ solutions were given by Nutku and Baekler \cite{Nutku:1989qi,Ortiz:1989vc}. Bianchi Type VIII-IX
exact solutions of TMG were studied in \cite{Aliev:1995cf} (using the spinor formulation). Supersymmetric solutions of the theory were found in \cite{Dereli:2001kq, Olmez:2005by}. In the massive double copy context, generalized shock wave solutions in the presence of spinning sources with specific choices were found in \cite{Gonzalez:2021ztm}. Similarly, exact solutions of NMG were constructed in \cite{Ahmedov:2010em, Ayon-Beato:2009cgh, Gurses:2011fv,Gurses:2015zia}. On the other hand, general Kundt solutions of TMG were investigated in \cite{Chow:2009vt}, in which explicit solutions with constant scalar invariants (CSI) were found. Kundt spacetimes CSI as well as non-CSI solutions of TMG and NMG were later constructed \cite{Chakhad:2009em}. \footnote{Note also that Kundt solutions of minimal massive gravity (MMG) \cite{Bergshoeff:2014pca}, which is another massive gravity theory in ${2+1}$ dimensions obtained from an action in the first-order formulation, were studied in \cite{Deger:2015wpa,Altas:2015dfa}}. The general Robinson-Trautman and Kundt solutions of 3D GR in the presence of pure radiation and gyratons have been discovered in \cite{Podolsky:2018zha}. However, to the best of our knowledge, there has been no systematic study of solutions of 3D massive theories of gravity in the presence of pure radiation as well as gyratons, which is the aim of this paper.

The layout of the paper is as follows: In Sec.~\ref{sc:RTKundt} we introduce the Robinson-Trautman and Kundt spacetimes by means of the existence of null geodesic congruence with non-vanishing and vanishing expansions. Then, in Sec.~\ref{eq:MG} we briefly review the main properties of TMG and NMG. The main part of the paper is contained in Sec.~\ref{sc:RT} and Sec.~\ref{sc:Kundt}, where we reduce the field equations for the Robinson-Trautman class with expansion ${\Theta=1/r}$ and the Kundt class ($\Theta=0$) for generic sources describing pure radiation and gyratons. Discussion for TMG is completely general. In the case of NMG, however, we also assume that $g_{ux}$ is quadratic in $r$, which is enforced automatically in TMG and 3D GR but not necessarily in NMG. We also study various subclasses and list a few solutions for specific sources. Finally, the paper is concluded with a brief discussion of our results in Sec.~\ref{sc:conclusions}.


\section{Robinson-Trautman and Kundt in 3D}\label{sc:RTKundt}

Let us first briefly review some general properties of the Robinson-Trautman and Kundt spacetimes in ${2+1}$ dimensions. In what follows, we will work in mostly positive signature $(-,+,+)$, use the geometric units in which ${c=G=1}$, and adopt the standard convention for the Riemann tensor, ${\left[\nabla_{\mu}, \nabla_{\nu}\right] V^{\kappa}=R^{\kappa}{}_{\lambda \mu\nu} V^{\lambda}}$.

It is well known \cite{Chow:2009vt,Podolsky:2018zha} that, under a very mild (and possibly removable) assumption that the spacetime admits a \textit{null geodesic congruence} generated by the null vector field $\boldsymbol{k}$, all geometries in ${2+1}$ dimensions are either Robinson-Trautman or Kundt. This follows from the fact that the optical matrix is one-dimensional, so its symmetric and anti-symmetric parts, which define \textit{twist} and \textit{shear}, trivially vanish. In addition to that, the metric of a generic Robinson-Trautman or Kundt spacetimes can be written in the \textit{canonical coordinates} $(r,u,x)$,
\begin{equation}\label{eq:RTKundt}
    \mathrm{d}s^2=g_{xx}(r,u,x)\mathrm{d}x^2+2g_{ux}(r,u,x)\mathrm{d}u \mathrm{d}x-2\mathrm{d}u \mathrm{d}r+g_{uu}(r,u,x)\mathrm{d}u^2 .
\end{equation}
Here, $u$ is a coordinate that foliates the spacetime by null hypersurfaces ${u=\text{const.}}$ to which $\boldsymbol{k}$ is orthogonal.\footnote{In contrast to higher dimensions, in 3D, any null vector field is geodesic if and only if it is hyper-surface orthogonal.} The coordinate $r$ is an affine parameter along the null congruence, ${\boldsymbol{k}=\boldsymbol{\partial}_r}$, and $x$ is a spatial coordinate, which is constant along $\boldsymbol{k}$, and span the one-dimensional transverse subspace of constant $u$ and $r$. In the canonical coordinates \eqref{eq:RTKundt}, the non-vanishing contravariant components of the metric tensor satisfy
\begin{equation}
    g^{ru}=-1,\quad g^{xx}=\frac{1}{g_{xx}}, \quad g^{rr}=-g_{uu}+g^{xx}g_{ux}^2, \quad g^{rx}=g^{xx}g_{ux}.
\end{equation}
Since $\boldsymbol{k}$ is trivially shear-free and twist-free, the only possibly nonzero optical scalar is the \textit{expansion} $\Theta=\partial_r g_{xx}/( 2g_{xx})$. The case of nonzero expansion ${\Theta \neq 0}$ defines the \textit{(expanding) Robinson-Trautman class} of geometries while the vanishing expansion ${\Theta =0}$ corresponds to the \textit{Kundt class} of geometries. 

For the sake of simplicity, we will focus only on the subclass of Robinson-Trautman spacetimes with the expansion ${\Theta={1}/{r}}$. We will refer to it as the \textit{`GR-like' Robinson-Trautman} because such expansion is automatically enforced by the field equations of 3D GR.\footnote{For example, when imposing $\Theta={1}/{r}$ in the $rr$ component of the field equation of TMG [see \eqref{eq:TMGfeq} below], 
\begin{equation*}
\sigma \left(\Theta^2+\partial_r \Theta\right)+\tfrac{\sqrt{g^{xx}}}{2\mu }\left[\partial_r^3 g_{ux}-2g_{ux}\partial_r^2 \Theta -4(\Theta g_{ux}+\partial_r g_{ux})\partial_r \Theta-4\Theta (\partial_x \Theta +\Theta \partial_r g_{ux})-2\partial_r\partial_x\Theta\right]=0 \to \frac{1}{\mu }\partial^3_r g_{ux}=0\;,
\end{equation*}
the GR term drops out while the rest gets drastically simplified allowing us to integrate the field equations.
} The field equations of TMG and NMG may in principle also admit also other non-zero expansions, however, the analysis of the special subcase ${\Theta=1/r}$ leads to nice polynomial dependencies of the metric functions in $r$ and allows for great simplification of the field equations. Specifically, we can introduce the function $P(u,x)$ such that
\begin{equation}
    g_{xx}=\frac{r^2}{P^2(u,x)}\;,
\end{equation}
and write the GR-like Robinson-Trautman metrics in the form
\begin{equation}\label{eq:RTmetric1overrexp}
    \mathrm{d}s^2=\frac{r^2}{P^2(u,x)}\mathrm{d}x^2+2g_{ux}(r,u,x)\mathrm{d}u \mathrm{d}x-2\mathrm{d}u \mathrm{d}r+g_{uu}(r,u,x)\mathrm{d}u^2 .
\end{equation}

In the Kundt class, the vanishing expansion, ${\Theta=0}$, implies that the spatial metric component $g_{xx}$ is independent of the coordinate $r$. Hence, we have
\begin{equation}
    g_{xx}=\frac{1}{P^2(u,x)}\;.
\end{equation}
Nevertheless, without loss of generality, we can set ${P=1}$ since by applying the coordinate  transformation ${x\to x'}$,
\begin{equation}
    x=\int P(u,x')\mathrm{d}x'
\end{equation} 
together with the re-definition
\begin{equation}
    P g_{ux} \to g_{ux}\;,
\end{equation}
the metric of a general Kundt spacetime will take the form
\begin{equation}\label{eq:KundtmetricP1}
    \mathrm{d}s^2 =\mathrm{d}x^2+2g_{ux}(r,u,x)\,\mathrm{d}u\mathrm{d}x-2\mathrm{d}u \mathrm{d}r+g_{uu}(r,u,x)\,\mathrm{d}u^2,
\end{equation}
where we dropped the primes labelling the new coordinate.


\section{3D massive gravity theories} \label{eq:MG}


\subsection{Topologically massive gravity (TMG)}

The Lagrangian density of TMG is given by \cite{Deser:1982vy}
\begin{equation}
{\cal L}= \sqrt{-g} \, \left[ \sigma R-2\Lambda +\frac{1}{2 \mu} \, \eta^{\mu \nu \alpha} \Gamma^\beta{_{\mu \sigma}} \Big (\partial_\nu \Gamma^\sigma{_{\alpha \beta}}+\frac{2}{3} \Gamma^\sigma{_{\nu \lambda}}  \Gamma^\lambda{_{\alpha \beta}} \Big ) \right],
\end{equation}
where the terms with the Christoffel symbols $\Gamma^\beta{_{\mu \sigma}}$ correspond to the \textit{gravitational Chern-Simons term}, $\mu$ is a mass parameter, ${\sigma=\pm 1}$ is a sign reversal parameter which depends on the unitarity requirements \footnote{Here, $\sigma=+1$ refers to ''correct'' sign while $\sigma=-1$ refers to ''wrong'' sign for the Einstein Hilbert term.}, $\Lambda$ is the cosmological constant, and $\eta^{\mu \nu \alpha} $ is the rank-3 tensor described by Levi-Civita symbol as $\epsilon^{\mu \nu \alpha}/\sqrt{-g}$. TMG is a parity non-invariant theory and describes massive spin-$2$ excitation with mass
\begin{equation}
    m_g^2=\mu^2\sigma^2+\Lambda.
\end{equation} 
Note that $\mu \to -\mu$ reverses the helicity of the graviton, but it keeps the mass invariant. On the other hand, in the flat space limit ($\Lambda \to 0$), the theory possesses a single massive degree of freedom with mass $m_g=\lvert  \mu \sigma  \rvert$ and $\sigma$ must be chosen negative for the positivity of kinetic energy (for the unitarity). Canonical analysis of TMG (and NMG) both in flat and (A)dS backgrounds was carried out in \cite{Gullu:2010sd} 
where masses and the correct signs were identified. When coupled to a source, the field equations of TMG are   
\begin{equation}\label{eq:TMGfeq}
\mathcal{E}^{\text{TMG}}_{\mu\nu}\equiv\sigma G_{\mu\nu}+\Lambda g_{\mu\nu}+\frac{1}{\mu} C_{\mu\nu}=8\pi T_{\mu\nu} ,
\end{equation}
where ${G_{\mu\nu}=R_{\mu\nu}-(1/2)R g_{\mu\nu}}$ is the usual Einstein tensor and $ C_{\mu \nu} $ is the symmetric, traceless and covariantly conserved Cotton tensor defined as 
\begin{equation}
\begin{aligned}
 C^{\mu \nu} &=\eta^{\mu \alpha \beta} \nabla_\alpha \Big ( R^\nu{_\beta}-\frac{1}{4} \delta^\nu{_\beta} R \Big )
 \\
 &=\frac{1}{2} \eta^{\mu \alpha \beta} \nabla_\alpha G^\nu{_\beta}+\frac{1}{2} \eta^{\nu \alpha \beta} \nabla_\alpha G^\mu{_\beta}.
\end{aligned}
\end{equation} 
In 3D, the Cotton tensor takes the role of the Weyl tensor as it is zero for conformally flat metrics. The Weyl tensor, on the other hand, identically vanishes. The field equations \eqref{eq:TMGfeq} are of the third order in derivatives of the metric. By setting ${\sigma=1}$ and taking the limit of infinite mass $\mu\to\infty$, we recover 3D GR. We will refer to this limit as the \textit{GR limit}.


\subsection{New massive gravity (NMG)}

 The Lagrangian density of NMG reads \cite{Bergshoeff:2009hq}
\begin{equation}
{\cal L}= \sqrt{-g} \, \left[ \sigma R-2\Lambda_0 +\frac{1}{m^2}\left(R_{\mu\nu}R^{\mu\nu}-\frac{3}{8}R^2\right)\right] ,
\end{equation}
in which $m$ is a mass parameter and $\Lambda_0$ is the \textit{bare} cosmological constant which is generally not equal to the \textit{effective} cosmological constant $\Lambda$. The theory has a massive graviton with a two graviton polarization about flat and (A)dS vacua, leading to NMG being a parity preserving theory, unlike the TMG. After coupling to the source, the field equations of NMG read 
\begin{equation}\label{eq:NMGfeq}
    \mathcal{E}^{\text{NMG}}_{\mu\nu}\equiv \sigma G_{\mu\nu} + \Lambda_0 g_{\mu\nu}+\frac{1}{2m^2}K_{\mu\nu} = 8\pi T_{\mu\nu}, 
\end{equation}
which are of the fourth order in derivatives of the metric. The covariantly conserved tensor $K_{\mu\nu}$ is defined as 
 \begin{equation}
     K_{\mu\nu} = 2\square R_{\mu\nu}-\frac12\nabla_{\mu}\nabla_{\mu}R-\frac12g_{\mu\nu}\square R+4R_{\mu\alpha\nu\beta}R^{\alpha\beta}-\frac32RR_{\mu\nu}-g_{\mu\nu}K\;.
  \end{equation}
Here, $K$ is the trace of $K_{\mu\nu}$ tensor defined as $K=g^{\mu\nu}K_{\mu\nu}=R_{\mu\nu}R^{\mu\nu}-(3/8)R^2$ which does not contain derivatives of the curvature. This property leads to NMG being free of a scalar ghost-like degree of freedom. On the other hand, maximally symmetric two-vacuum solutions can be given as
\begin{equation}\label{eq:Lambdapm}
    \Lambda_{\pm}=-2\sigma m^2\pm 2\sqrt{\Lambda_0 m^2+m^4},
\end{equation}
as long as ${\Lambda_0 \geq -m^2}$. When this bound is saturated, a unique vacuum exists: anti-de Sitter for positive $\sigma$ and de Sitter for negative $\sigma$. Similar to the TMG case, the GR limit corresponds to choosing ${\sigma=1}$ and taking the limit of infinite mass ${m\to\infty}$.


\section{GR-like Robinson-Trautman gyratons in TMG \& NMG} \label{sc:RT}

In this section, we consider the GR-like Robinson-Trautman metrics given by \eqref{eq:RTmetric1overrexp}. We look for solutions of TMG and NMG with the only non-vanishing components of the energy-momentum tensor $T_{\mu\nu}$ being the pure radiation component $T_{uu}$ and the gyratonic matter component $T_{ux}$. The trace of the energy-momentum tensor vanishes, ${T=0}$. Since the energy-momentum is conserved (in TMG as well as NMG), ${\nabla^\mu T_{\mu\nu}=0}$, we have the following constraints
\begin{equation}
\begin{aligned}
   & \partial_r T_{ux}+\frac{1}{r} T_{ux}=0,\\&
    \partial_r T_{uu}+\frac{1}{r} T_{uu}=\frac{P^2}{r^2}\left[\partial_x T_{ux}+\left(\partial_r g_{ux}-\frac{2 g_{ux}}{r}+\frac{\partial_x P}{P}\right)T_{ux}\right].
    \label{EMTconsRT}
    \end{aligned}
\end{equation}


\subsection{Solving TMG}

Let us first impose the TMG field equations \eqref{eq:TMGfeq} and integrate the non-trivial equations corresponding to the non-vanishing components of $\mathcal{E}^{\text{TMG}}_{\mu\nu}$ one by one.
\begin{enumerate}
    \item\underline{$\mathcal{E}^{\text{TMG}}_{rr}=0$} \newline
In this case, the $rr$ component of the field equation turns out to be
\begin{equation}
    \frac{1}{\mu }\partial^3_r g_{ux}=0.
\end{equation}
Thus, $g_{ux}$ must take the form
\begin{equation}\label{eq:guxTMG}
    g_{ux}= c_1+c_2 r+c_3 r^2,
\end{equation}
in which $c_1$, $c_2$, and $c_3$ are arbitrary functions of coordinates $u$ and $x$. 
\item \underline{$\mathcal{E}^{\text{TMG}}_{rx}=0$} \newline
The $rx$ component of the field equation becomes
\begin{equation}
 \frac{\sigma c_2}{2r}+\frac{r \partial^3_r g_{uu}}{4\mu P}=0
\end{equation}
and yields a general solution
\begin{equation}
    g_{uu}= c_4+c_5 r+c_6 r^2+2\mu \sigma c_2 r (\log r-1) P ,
\end{equation}
where $c_4$, $c_5$, and $c_6$ are again arbitrary functions of coordinates $u$ and $x$. At this stage, it proves useful to consider the trace of the field equations, ${R={6\sigma\Lambda}}$, which provides several relations among the functions $c_i$ and $P$. After expressing the Ricci scalar of our metric in terms of the functions $c_i$ and $P$,
\begin{equation}
\begin{aligned}
    R= \frac{1}{2r^3}\bigg[&(12c_6-12P^2c_3^2)r^3+\left(4c_5+4\sigma\mu P c_2-8Pc_3\partial_x P-8P^2c_2c_3-8P^2\partial_x c_3-\frac{8\partial_x P}{P}\right)r^2\\&+(-4P\partial_x P c_2+P^2c_2^2-4P^2\partial_x c_2)r+4P^2c_1c_2+8\sigma\mu Pc_2 r^2 \log r\bigg],
    \end{aligned}
\end{equation}
and demanding ${R={6\Lambda}/{\sigma}}$ to hold for every value of $r$, we can find that
\begin{equation}
  c_2=0,\quad c_6= \frac{\Lambda}{\sigma}+c_3^2 P^2 ,\quad c_5= 2c_3 P\partial_x P+2 P^2\partial_x c_3+2\frac{\partial_u P}{P}.
\end{equation}
Furthermore, without loss of generality, we can set ${c_3=0}$ by means of the coordinate transformation ${x\to x'}$,
\begin{equation}\label{eq:gaugetr}
    x=-\int P^2(u,x')\,c_3(u,x') \mathrm{d}u,
\end{equation}
followed by the re-definitions
\begin{equation}\label{eq:redef}
    \frac{P}{\frac{\partial x}{\partial x'}} \to P, \quad c_1 \frac{\partial x}{\partial x'}\to c_1, \quad c_4 -2 P^2 c_3\,c_1\to c_4.
\end{equation}
Hence, after dropping the primes, our metric can be transformed into the form
\begin{equation}\label{eq:newmet}
    \mathrm{d}s^2=\frac{r^2}{P^2}\mathrm{d}x^2+2c_1 \mathrm{d}u \mathrm{d}x-2\mathrm{d}u \mathrm{d}r+\left(c_4+2r\partial_u \log P +\frac{\Lambda}{\sigma} r^2\right)\mathrm{d}u^2.
\end{equation}
\item \underline{$\mathcal{E}^{\text{TMG}}_{ru}=0$ and $\mathcal{E}^{\text{TMG}}_{xx}=0$}  \newline
The $ru$ and $xx$ components of the field equations yield a single constraint
\begin{equation}
     \frac{1}{\mu }\left[2c_1\partial_u P+P(\partial_x c_4-2\partial_u c_1)\right]=0.
    \label{rucompeq}
\end{equation}
\item \underline{$\mathcal{E}^{\text{TMG}}_{ux}=8\pi T_{ux}$} \newline
Before giving field equation, let us first consider the energy-momentum conservation equations \eqref{EMTconsRT}. They can be integrated to determine the gyratonic matter and pure radiation components of the energy-momentum tensor,
\begin{equation}
    \begin{aligned}
       & T_{ux}=\frac{J}{r},\\&
       T_{uu}=\frac{N}{r}+\frac{P^2J\,c_1}{r^3}-\frac{P}{r^2}\partial_x(PJ),
    \end{aligned}
\end{equation}
where the gyratonic and pure radiation \textit{profile functions} $J$ and $N$ are arbitrary functions of $u$ and $x$. With these definitions of profile functions, the $ux$ component of the field equation gives rise to the equation for $J$
\begin{equation} \label{1.consRT}
\begin{aligned}
 &-\frac{1}{\sigma\mu P^2} \Big[2\Lambda P^3 \partial_x c_1+2\sigma \left(-c_4+(\partial_x P)^2\right) \partial_u P+\sigma P \left(2(-\sigma\mu c_1-\partial_x^2 P)\partial_u P+\partial_u c_4-2 \partial_x P \partial_u \partial_x P \right)\\&
 +P^2 \left(2\Lambda c_1 \partial_x P+ \left(-\mu  (\partial_x c_4-2\partial_u c_1)+2\sigma\partial_u \partial^2_x P\right)  \right) \Big]=16\pi  J.
  \end{aligned}
\end{equation}
\item \underline{$\mathcal{E}^{\text{TMG}}_{uu}=8\pi T_{uu}$} \newline
Similarly, the $uu$ component of the TMG field equations leads to the equation with the profile function $N$ on the right-hand side,
\begin{equation}
    -2\Lambda P^2\partial_x c_1+2\sigma \left(c_4-(\partial_x P)^2\right)\partial_u \log P+\sigma \big(2\partial_x(\partial_x P\partial_u P)-\partial_u c_4\big)-2P(\Lambda c_1 \partial_x P+\sigma \partial_u \partial^2_x P)=16\pi  N,
    \label{2.consRT1}
\end{equation}
\end{enumerate}

In summary, the GR-like Robinson-Trautman solution of TMG can be written in the form [see \eqref{eq:newmet}]
\begin{equation}\label{eq:GRlikeRTTMG}
   \mathrm{d}s^2 =\frac{r^2}{P^2}\mathrm{d}x^2+2c_1\,\mathrm{d}u \mathrm{d}x-2\mathrm{d}u \mathrm{d}r+\left(c_4+2r\partial_u \log P+\frac{\Lambda}{\sigma}r^2\right)\,\mathrm{d}u^2,
\end{equation}
where $P$, $c_1$, and $c_4$ are arbitrary functions of $u$ and $x$ satisfying \eqref{rucompeq}, \eqref{1.consRT}, and \eqref{2.consRT1}. These equations can be satisfied as follows: If we prescribe an arbitrary $P$ and $c_1$, we can obtain $c_4$ by integrating \eqref{rucompeq}, which is linear in $c_4$. The equations \eqref{1.consRT} and \eqref{2.consRT1} then determine the gyratonic profile function $J$ and pure radiation profile function $N$. It is also important to mention that the metric function $c_1$ remains arbitrary like in 3D GR. This property does not hold in dimensions higher than three. Note also that, in the GR limit, the constraint \eqref{rucompeq} is trivially satisfied since the left-hand side vanishes, and the equations \eqref{1.consRT} and \eqref{2.consRT1} reduce to
\begin{equation}
\begin{aligned}
    \sigma \partial_x c_4-2\sigma \partial_u c_1+2\sigma c_1 \partial_u \log P &= 16\pi  J,\\
    -2\Lambda P^2\partial_x c_1+2\sigma \left(c_4-(\partial_x P)^2\right)\partial_u \log P+\sigma \big(2\partial_x(\partial_x P\partial_u P)-\partial_u c_4\big)-2P(\Lambda c_1 \partial_x P+\sigma \partial_u \partial^2_x P)&=16\pi  N,
\end{aligned}
\end{equation}
which are the correct GR limits \cite{Podolsky:2018zha}.


\subsubsection{Subcase ${c_1=0}$}

In this case, due to \eqref{rucompeq}, the metric function $c_4$ is independent of the coordinate $x$, that is $c_4=c_4(u)$. Therefore, the metric \eqref{eq:GRlikeRTTMG} takes a simpler form,
\begin{equation}
    \mathrm{d}s^2=\frac{r^2}{P^2}\mathrm{d}x^2-2\mathrm{d}u\mathrm{d}r+\left(c_4+2r\partial_u \log P+\frac{\Lambda}{\sigma}r^2\right)\,\mathrm{d}u^2,
\end{equation}
and the remaining field equations reduce to
\begin{equation}
    \begin{aligned}
        -\frac{1}{\sigma\mu P^2} \left[2\sigma (-c_4+(\partial_x P)^2) \partial_u P+\sigma P (-2\partial_x( \partial_xP\partial_u P)+\partial_u c_4 )
 +2P^2 \sigma \partial_u \partial^2_x P   \right] &=16\pi J,\\
        2\sigma (c_4-(\partial_x P)^2)\partial_u \log P+\sigma (2\partial_x( \partial_x P\partial_u P)-\partial_u c_4)-2P\sigma \partial_u \partial^2_x P &=16\pi  N.
    \end{aligned}
\end{equation}
which give rise to ${N=\sigma\mu P^2 J}$. Observe that in the $J=0$ case which leads to $N=0$, all solutions are also solutions of $3D$ GR. To find non-Einstenian solutions, one needs to consider the non-zero gyratonic matter component $J$. Notice also that field equations are independent of cosmological constant $\Lambda$. They can be satisfied by prescribing arbitrary metric functions $c_4$ and $P$, for which we can always find the corresponding $N$ and $J$. Alternatively, if we prescribe $P$ together with either $J$ or $N$, we can solve the linear equation for $c_4$. 

If we also set $P=1$, the equations further reduce to
\begin{equation}
    \begin{aligned}
        -\frac{1}{\mu} \partial_u c_4 &=16\pi J,\\
       -\sigma  \partial_u c_4 &=16\pi  N ,
    \end{aligned}
\end{equation}
which leads to ${N=\sigma\mu J}$. The solution of such equations can be interpreted as a Vaidya-type object emitting pure radiation.

Observe that since the field equations are very complex, finding a general solution is a difficult task. Nevertheless, we can find some new solutions with some assumptions. In this respect, to give an explicit solution in the vacuum, let us assume that $c_4=$ constant and $P(u,x)=\tilde{c}_1 u+\tilde{c}_2 x$ (which leads to $c_4={\tilde{c}}_2^2$). In this case, the special solution is
\begin{equation}
     \mathrm{d}s^2=\frac{r^2}{(\tilde{c}_1 u+\tilde{c}_2 x)^2}\mathrm{d}x^2-2\mathrm{d}u\mathrm{d}r+\left(\tilde{c}_2^2+\frac{2\tilde{c}_1}{\tilde{c}_1 u+\tilde{c}_2 x}\,r+\frac{\Lambda}{\sigma}r^2 \right)\mathrm{d}u^2,
\end{equation}
where $\tilde{c}_1$ and $\tilde{c}_2$ are real constants. 

\subsection{Solving NMG}

Let us now move on to the field equations of NMG \eqref{eq:NMGfeq}. Again, we will go through the non-trivial equations corresponding to non-zero $\mathcal{E}^{\text{NMG}}_{\mu\nu}$ one by one.
\begin{enumerate}
    \item \underline{$\mathcal{E}^{\text{NMG}}_{rr}=0$}\newline
The $rr$ component of the field equation leads to the constraint
\begin{equation}
\begin{aligned}
    &\frac{1}{m^2}\Big[-2r^2 P \partial_x P \partial_r^3 g_{ux}+r^3(2\partial_r^3 g_{uu}+r\partial_r^4 g_{uu})+P^2\big((\partial_r g_{ux})^2-r \partial_r g_{ux}(2 \partial_r^2 g_{ux} +3r \partial_r^3 g_{ux})\\&+2 g_{ux}r\partial_r^3 g_{ux}+r^2((\partial_r^2 g_{ux})^2-2(\partial_r^3\partial_x g_{ux}+g_{ux}\partial_r^4 g_{ux}))\big)\Big]=0.
    \end{aligned}
\end{equation}
In order to solve this equation, we need to consider an \textit{additional assumption}. We will assume that $g_{ux}$ is again a polynomial function in $r$,
\begin{equation}
    g_{ux}= c_1+c_2 r+c_3 r^2,
\end{equation}
where $c_1$, $c_2$, and $c_3$ are arbitrary functions of coordinates $u$ and $x$. This form of $g_{ux}$ is motivated by the 3D GR and TMG (cf. \eqref{eq:guxTMG}), for which it is actually enforced by the field equation. In the case of NMG, however, it may not be the most general form of $g_{ux}$ for our ansatz \eqref{eq:RTmetric1overrexp}. Nevertheless, considering this assumption significantly simplify further calculations. The $rr$ component of the field equation then takes a manageable form,
\begin{equation}
    P^2 c_2^2+2r^3 \partial_r^3 g_{uu}+ r^4 \partial_r^4 g_{uu}=0,
\end{equation}
which yields
\begin{equation}
   g_{uu}= c_4+c_5 r+ c_6 r^2+\frac{1}{2}P^2 c_2^2\, \log r-c_7r\log r,
\end{equation}
where $c_4$, $c_5$, $c_6$, and $c_7$ are again some functions of coordinates $u$ and $x$. It turns out that, without loss of generality, we can set ${c_3=0}$ by the coordinate transformation, ${x\to x'}$,
\begin{equation}\label{eq:xPc3du}
    x=-\int P^2 (u,x')\,c_3 (u,x') \mathrm{d}u,
\end{equation}
and re-definitions 
\begin{equation}\label{eq:PtoP}
    \frac{P}{\frac{\partial x}{\partial x'}} \to P, \quad c_1 \frac{\partial x}{\partial x'}\to c_1, \quad c_2 \frac{\partial x}{\partial x'}\to c_2,  \quad c_4-2 P^2 c_3\,c_1\to c_4, \quad c_5-2 P^2 c_3\,c_2\to c_5, \quad c_6- P^2 c_3^2\to c_6.
\end{equation}
Our metric then transforms into the form
\begin{equation}
    \mathrm{d}s^2=\frac{r^2}{P^2}\mathrm{d}x^2+2(c_1+ c_2 r) \mathrm{d}u \mathrm{d}x-2\mathrm{d}u \mathrm{d}r+\left(c_4+c_5 r+c_6 r^2+\tfrac{1}{2}P^2 c_2^2\, \log r-c_7r\log r  \right)\mathrm{d}u^2.
\end{equation}
\item \underline{$\mathcal{E}^{\text{NMG}}_{rx}=0$} \newline
The $rx$ component of the field equation yields
\begin{equation}
\begin{aligned}
 &24 P^2 c_1 c_2^2+\left(15P^2c_2^3-22Pc_2^2\partial_x P-12c_2\,c_4-22 c_2 P^2\partial_x c_2+12 c_1\,c_5+10c_1\,c_7+12\partial_x c_4-24\partial_u c_1\right)r\\&+\left(4 \partial_x c_7-2c_2\,c_7\right)r^2+\left(2c_2(c_6+2\sigma m^2)\right)r^3+ \left(-6P^2c_2^3+12Pc_2^2\partial_x P+12c_2 P^2\partial_x c_2-12 c_1\, c_7\right)r \log r=0.
 \label{rxnmg}
  \end{aligned}
\end{equation}
By equating the coefficients in front of the $r$ dependent terms to zero, we find that\footnote{Specifically, there are 3 possibilities to satisfy \eqref{rxnmg}: i) ${c_1=0}$, ${c_2 \neq 0}$, ii) ${c_1\neq 0}$, ${c_2 =0}$, iii) ${c_1=c_2=0}$. The trace of the field equations for i) leads to ${c_2=0}$ which invalidates this case. Hence, we can consider either ii) or iii). In what follows, we focus on ii) since the case iii) actually leads to the same as setting ${c_1=0}$ (with ${c_4=c_4(u)}$) in \eqref{1.consRTNMG}, \eqref{2.consRTNMG} and \eqref{nmgsol} later on.}
\begin{equation}
    c_2=0, \quad c_7=0, \quad c_5=\frac{2\partial_u c_1-\partial_x c_4}{c_1}.
    \label{condnmg}
\end{equation}

\item \underline{$\mathcal{E}^{\text{NMG}}_{xx}=0$ and $\mathcal{E}^{\text{NMG}}_{ru}=0$}

The equations for $xx$ and $ru$ components lead to 
\begin{equation}
\begin{aligned}
    &\frac{1}{m^2}\Big[-4 \partial_x P
   P^2 c_1 \left(2 c_6c_1^3-(\partial_x^2c_4-2
   \partial_x\partial_u c_1)
   c_1+\partial_x c_1\left(\partial_x c_4-2\partial_u c_1\right)\right)-4 P^3 \Big(3 \partial_x c_6 c_1^4+2 c_6
   \partial_x c_1 c_1^3\\&-\left(\partial_x^3 c_4-2\partial_u\partial_x^2 c_1\right)
   c_1^2+c_1\left(\partial_x^2 c_1\left(\partial_x c_4-2\partial_u c_1\right)+2
   \partial_x c_1\left(\partial_x^2 c_4-2 \partial_u\partial_x c_1\right)\right)
    -2\left(\partial_x c_1\right)^2\left(\partial_x c_4-2
   \partial_u c_1\right)\Big)\\&+P c_1^3 \left(r^3 \left(4
   \Lambda_0 m^2-c_6(c_6+4 m^2 \sigma\right)\right)-4
   \partial_u c_4)+8
   \partial_u P c_1^3 c_4 \Big]=0
   \label{xxnmg0}
   \end{aligned}
\end{equation}
and
\begin{equation}
\begin{aligned}
\frac{1}{m^2}&\Big[-2 \partial_u P c_1^3 \left(r^2
   \left(c_6+2 m^2 \sigma\right)-2 c_4\right)+Pc_1^2 \Big(c_1 \left(r^3
   \left(4 m^2 \sigma c_6+c_6^2-4 \Lambda_0 m^2\right)-2
   \partial_u c_4\right)\\&-2 \partial_x P
   P^2 c_1\left(2 c_6 c_1^3-\left(\partial_x^2 c_4-2\partial_u\partial_x c_1\right)c_1+\partial_x c_1\left(\partial_x c_4-2 \partial_u c_1\right)\right)-2 P^3 \Big(3 \partial_x c_6 c_1^4+2 c_6
   \partial_x c_1 c_1^3\\&-\left(\partial_x^3 c_4-2
   \partial_u\partial_x^2c_1\right)c_1^2+\left(\partial_x^2 c_1
   \left(\partial_x c_4-2 \partial_u c_1\right)+2\partial_x c_1
   \left(\partial_x^2 c_4-2 \partial_u\partial_x c_1\right)\right) 
   c_1-2\left(\partial_x c_1\right)^2\left(\partial_x c_4-2
   \partial_u c_1\right)\Big)\\&-r^2\left(\partial_x c_4-2
   \partial_u c_1\right) \left(c_6+2m^2 \sigma\right)\Big)\Big]=0.
   \label{runmg}
\end{aligned}
\end{equation}
After combining these two equations into ${\mathcal{E}^{\text{NMG}}_{xx}-2\mathcal{E}^{\text{NMG}}_{ru}}$, we get
\begin{equation}
   r\left( 12 m^2 \Lambda_0-3c_6(4\sigma m^2+c_6)\right)+\frac{4(2\sigma m^2+c_6)\partial_u P}{P}+\frac{2(2\sigma m^2+c_6)(\partial_x c_4-2\partial_u c_1)}{c_1}=0,
   \label{trfeqNMG}
\end{equation}
which is actually just the trace of the field equations. From \eqref{trfeqNMG}, it is easy to see that
\begin{equation}
    c_6=-2\sigma m^2\pm 2\sqrt{\Lambda_0 m^2+m^4},
    \label{c6ef}
\end{equation}
meaning that $c_6$ is actually equal to effective cosmological constant $\Lambda_{\pm}$ for the maximally symmetric vacuum solutions, see \eqref{eq:Lambdapm}. Consequently, \eqref{trfeqNMG} reduces to
\begin{equation}
   \frac{1}{m^2}\left[2 c_1 \partial_u P+P (\partial_x c_4-2\partial_u c_1)\right]=0.
   \label{TraceNMG}
\end{equation}
Notice that this is the same constraint as the one we obtained for TMG, cf. \eqref{rucompeq}. Using \eqref{c6ef}
and \eqref{TraceNMG}, one can actually see that \eqref{xxnmg0} and \eqref{runmg} turn into a single equation
\begin{equation}
\begin{aligned}
    &\frac{1}{m^2}\Big[8 c_1^3 c_4 \partial_u P-4 P c_1^3 \partial_u c_4-4 \partial_x P
   P^2 c_1 \left(2 c_6c_1^3-\left(\partial_x^2c_4-2
   \partial_x\partial_u c_1\right)
   c_1+\partial_x c_1\left(\partial_x c_4-2\partial_u c_1\right)\right)-4 P^3 \Big(2 c_6
   \partial_x c_1 c_1^3\\&-\left(\partial_x^3 c_4-2\partial_u\partial_x^2 c_1\right)
   c_1^2+c_1\left(\partial_x^2 c_1\left(\partial_x c_4-2\partial_u c_1\right)+2
   \partial_x c_1\left(\partial_x^2 c_4-2 \partial_u\partial_x c_1\right)\right)
    -2\left(\partial_x c_1\right)^2\left(\partial_x c_4-2
   \partial_u c_1\right)\Big)\Big]=0.
   \label{xxnmg}
   \end{aligned}
\end{equation}

\item \underline{$\mathcal{E}^{\text{NMG}}_{ux}=8\pi T_{ux}$}\newline
Again, we can introduce profile functions $J$ and $N$ by integrating the energy-momentum conservation equation \eqref{EMTconsRT},
\begin{equation}
    \begin{aligned}
       & T_{ux}=\frac{J}{r},\\&
       T_{uu}=\frac{N}{r}+\frac{P^2J\,c_1}{r^3}-\frac{P}{r^2}\partial_x(PJ).
    \end{aligned}
\end{equation}
The $ux$ component of the field equation now reads
\begin{equation}
\begin{aligned}
 &\frac{1}{m^2r^2 P^2} \big\{8 P^4 c_6(-c_1\partial_x c_1+r\partial_x^2c_1)+8r \partial_u P\left[c_4\partial_x P-(\partial_x P)^3+3c_1 \partial_u P\right]+8P^3\big[-c_1^2c_6\partial_x P+c_1(rc_6\partial_x^2 P-\partial_u\partial_x^2 P)\\&+r(3c_6\partial_x P\partial_x c_1+\partial_u\partial_x^3P)\big] -8P\big[r\partial_u P (-2\partial_x P \partial_x^2 P+3 \partial_u c_1)+r(c_4-(\partial_x P)^2)\partial_u\partial_x P+c_1 ((-c_4+(\partial_x P)^2)\partial_u P
 \\
 &+r\partial_u^2 P)\big]+P^2 \big[c_1(8 rc_6(\partial_x P)^2+8 \partial_x(\partial_x P\partial_u P)-4 \partial_u c_4)+8r(-\partial_x^3 P\partial_u P
 -2 \partial_x^2P \partial_u\partial_x P+\partial_u^2 c_1)\big]\big\}=32\pi  J.
  \label{1.consRTNMG}
  \end{aligned}
\end{equation}

\item \underline{$\mathcal{E}^{\text{NMG}}_{uu}=8\pi T_{uu}$} \newline
The $uu$ component of the field equation yields
\begin{equation}
\begin{aligned}
 &\frac{1}{ m^2 r^3 P} \bigg[ P^4 \left(-c_1 \left(c_6\left(3  \partial_x P
   \partial_x c_1-r\partial_x^3 P\right)+\partial_x^3\partial_u P\right)+r \left(c_6 \left(4
    \partial_x^2 P\partial_x c_1+6\partial_x P \partial_x^2 c_1\right )+\partial_x^4\partial_u P\right)-c_6 c_1^2 \partial_x^2 P\right)\\&+P^3(c_1 \left(2 \partial_x^2  P
   \left(2 r  \partial_x P c_6+\partial_x\partial_u P\right)-\partial_u^2 c_1+\partial_x^3 P
   \partial_u P\right)+r (7(\partial_x P)^2 c_6\partial_x c_1+\partial_x\partial_u^2 c_1
   +2 \partial_x^3\partial_u P\partial_x P-\partial_x^4 P \partial_u P\\&-3 \partial_x^3 P
   \partial_x\partial_u P-2 \partial_x^2  P \partial_x^2\partial_u P)-c_1^2 c_6( \partial_x P)^2 )+P^2(c_1 (r (\partial_x P)^3c_6+\partial_u P (3
   \partial_u c_1-2  \partial_x P\partial_x^2  P)+\partial_x\partial_u P c_4\\&-r
   \partial_x\partial_u^2 P-\partial_x\partial_u P (  \partial_x P)^2)+r
   ( \partial_x P(\partial_u^2c_1+\partial_x^3 P \partial_u P)-5
   \partial_x\partial_u P \partial_u c_1-3\partial_u P \partial_x\partial_u c_1-\partial_x^2\partial_u P c_4-\partial_u^2 P\partial_x c_1\\&+\partial_x^2\partial_u P
   ( \partial_x P)^2+2 \partial_x^2  P \partial_x\partial_u P  \partial_x P+2 ( \partial_x^2  P)^2 \partial_u P)+\partial_u^2 P c_1^2)+P(\partial_u P (r(\partial_x^2 P\left(c_4-3
   ( \partial_x P)^2\right)+3 \partial_u P \partial_x c_1\\&+2\partial_x P \partial_u c_1)-3
   \partial_u P c_1^2+c_1\left(( \partial_x P)^3- \partial_x P c_4))+r
  \partial_x\partial_u P \left( \partial_x P c_4+8 \partial_u P
   c_1-( \partial_x P)^3\right)\right)+r  \partial_x P \partial_u P\\&
   \left(- \partial_x P c_4-5 \partial_u P c_1+( \partial_x P)^3\right)+P^5 c_6 \left(r
   \partial_x^3c_1-c_1 \partial_x^2 c_1\right)\bigg]=8\pi  N .
  \label{2.consRTNMG}
  \end{aligned}
\end{equation}
\end{enumerate}

To sum up, the GR-like Robinson-Trautman solutions of NMG in the presence of the source can be written as
\begin{equation}
   \mathrm{d}s^2 =\frac{r^2}{P^2}\mathrm{d}x^2+2c_1\,\mathrm{d}u \mathrm{d}x-2\mathrm{d}u \mathrm{d}r+(c_4+2r\partial_u \log P+c_6 r^2)\,\mathrm{d}u^2,
   \label{nmgsol}
\end{equation}
where $c_6$ is the effective cosmological constant given by \eqref{c6ef} and $P$, $c_1$, and $c_4$ are arbitrary functions of $u$ and $x$ satisfying \eqref{TraceNMG}, \eqref{xxnmg}, \eqref{1.consRTNMG} and \eqref{2.consRTNMG}. They can be solved by first prescribing $P$, and finding $c_1$ and $c_4$ from \eqref{TraceNMG} and \eqref{xxnmg}. The equations \eqref{1.consRTNMG} and \eqref{2.consRTNMG} then determine the gyratonic profile function $J$ and pure radiation profile function $N$, respectively. As in the case of 3D GR and TMG, the metric function $c_1$ remains arbitrary, unlike the case of dimensions higher than three. In the GR limit, \eqref{TraceNMG} and \eqref{xxnmg} are again automatically satisfied while \eqref{1.consRTNMG} and \eqref{2.consRTNMG} take the form
\begin{equation}
\begin{aligned}
\quad 2 c_1 \frac{\partial_u P}{P}+ (\partial_x c_4-2\partial_u c_1)&=16\pi  J,\\
-\Lambda P^2 \partial_x c_1+(c_4-(\partial_x P)^2)\partial_u \log P+\partial_u P \partial_x^2 P-\frac{1}{2}\partial_u c_4+\partial_x P\partial_u\partial_x P-P(\Lambda c_1 \partial_x P+\partial_u\partial_x^2 P)&=8\pi  N,
\end{aligned}
\end{equation}
which matches the 3D GR results from \cite{Podolsky:2018zha}.


\subsubsection{Subcase ${c_1=0}$}

In this case, due to  \eqref{rxnmg}, we have ${c_4=c_4(u)}$. Consequently, the metric \eqref{nmgsol} simplifies to
\begin{equation}
    \mathrm{d}s^2=\frac{r^2}{P^2}\mathrm{d}x^2-2\mathrm{d}u\mathrm{d}r+(c_4+2r\partial_u \log P+c_6 r^2)\mathrm{d}u^2
\end{equation}
and the only independent equations are
\begin{equation}
    \begin{aligned}
     &\frac{1}{m^2r P^2} \big\{8 \partial_u P\big[c_4\partial_x P-(\partial_x P)^3\big]-8P\big[-2\partial_u P \partial_x P \partial_x^2 P+(c_4-(\partial_x P)^2)\partial_u\partial_x P\big]+8P^2(-\partial_x^3 P\partial_u P-2\partial_x^2 P\partial_u\partial_x P)\\&\hskip 1.5 cm+8P^3\partial_u\partial_x^3P\big\}=32\pi J
    \end{aligned}
\end{equation}
and
\begin{equation}
    \begin{aligned}
     &\frac{1}{ m^2 r^2 P} \bigg[    P^2 \left(\partial_u\partial_x^2 P\left((\partial_x P)^2-c_4\right)+2 \partial_u P (\partial_x^2 P)^2+2 \partial_x P \partial_x\partial_u P
   \partial_x^2 P+\partial_x P \partial_x^3 P \partial_u P\right)+P( \partial_x^2 P \partial_u P\left(c_4-3 (\partial_x P)^2\right)\\&+ \partial_x P \partial_x\partial_u P\left(c_4-(\partial_x P)^2\right))+ (\partial_x P)^2 \partial_u P\left((\partial_x P)^2-c_4\right)+ \partial_x^4\partial_u P P^4-P^3(\partial_x^4 P \partial_u P+3 \partial_x^3 P \partial_x\partial_u P+2 \partial_x^2 P \partial_x^2\partial_u P\\&-2 \partial_x P \partial_x^3\partial_uP) \bigg] =8\pi  N.
    \end{aligned}
\end{equation}
Similar to the cases of 3D GR and TMG, the field equations do not depend on the bare cosmological constant $\Lambda_0$ (and hence $\Lambda$).

To give an explicit solution in the vacuum, let us assume that $P(u,x)=\tilde{c}_1x+\tilde{c}_2u$ which leads to $c_4=\tilde{c}_2^2$. In this case, the solution is
\begin{equation}
     \mathrm{d}s^2=\frac{r^2}{{(\tilde{c}_1 u+\tilde{c}_2x)}^2}\mathrm{d}x^2-2\mathrm{d}u\mathrm{d}r+\left(\tilde{c}_2^2+\frac{2\tilde{c}_1}{\tilde{c}_1u+\tilde{c}_2x}r+c_6 r^2\right)\mathrm{d}u^2
     \label{nmgexpsol}
\end{equation}
where $\tilde{c}_1$ and $\tilde{c}_2$ are real constants. The metric \eqref{nmgexpsol} is not solution for 3D GR.


\section{Kundt Gyratons in TMG \& NMG} \label{sc:Kundt}

Now let us consider a general Kundt spacetime \eqref{eq:KundtmetricP1} and look for solutions within TMG and NMG. As in the previous sections, we suppose that the only non-vanishing components of the energy-momentum tensor $T_{\mu\nu}$ are pure radiation component $T_{uu}$ and gyratonic matter component $T_{ux}$. The trace again vanishes, ${T=0}$. From the energy-momentum conservation, one can find the following constraints
\begin{equation}
\begin{aligned}
   & \partial_r T_{ux}=0,\\&
    \partial_r T_{uu}=(\partial_x+\partial_r g_{ux}) T_{ux}.
    \label{EMTcons}
    \end{aligned}
\end{equation}


\subsection{Solving TMG}

First, we analyze the source-coupled field equations of TMG.
\begin{enumerate}
    \item \underline{$\mathcal{E}^{\text{TMG}}_{rr}=0$ } \newline
The $rr$ component of the field equation gives 
\begin{equation}
    \partial^3_r g_{ux}=0,
\end{equation}
so $g_{ux}$ can is given by
\begin{equation}
    g_{ux}= c_1+c_2r+c_3r^2,
\end{equation}
where $c_1$, $c_2$, and $c_3$ are arbitrary functions of coordinates $u$ and $x$. 
\item \underline{$\mathcal{E}^{\text{TMG}}_{rx}=0$} \newline
The $rx$ component of field equation leads to 
\begin{equation}
    \partial^3_r g_{uu}=4\sigma\mu c_3,
\end{equation}
from which we can find that $g_{uu}$ takes the form
\begin{equation}
    g_{uu}= c_4+c_5r+c_6r^2+\frac{2}{3}\sigma\mu c_3r^3 ,
\end{equation}
where $c_4$, $c_5$ and $c_6$ are some arbitrary functions of $u$ and $x$. At this stage, it is useful to study the trace of field equations of TMG, ${R=6\sigma\Lambda}$. Since the Ricci scalar for the Kundt metric is given by
\begin{equation}
    R= -10 c^2_3 r^2+r (-10 c_2 c_3+4\sigma\mu c_3-4\partial_x c_3)+\left(-4c_1 c_3+2c_6-\frac{3}{2}c^2_2-2\partial_x c_2\right),
\end{equation}
we have 
\begin{equation}\label{eq:c6}
   c_3=0, \quad c_6= \frac{1}{4}\left(3c^2_2+4\partial_x c_2+\frac{12\Lambda}{\sigma}\right). 
\end{equation}

\item \underline{$\mathcal{E}^{\text{TMG}}_{ru}=0$ and $\mathcal{E}^{\text{TMG}}_{xx}=0$}  \newline
The $ru$ as well as the $xx$ component of field equations becomes
\begin{equation}
  \partial_x\left(\partial_x c_2+\frac{1}{2}c^2_2+\frac{2\Lambda}{\sigma}\right)+\left(-\sigma\mu+\frac{3}{2}c_2\right)\left(\partial_x c_2+\frac{1}{2}c^2_2+\frac{2\Lambda}{\sigma}\right)=0 ,
  \label{eqru}
\end{equation}
This equation can be solved for $c_2$, which then determines also $c_6$ through \eqref{eq:c6}. For instance, this equation is satisfied if ${c_2=2\sigma\mu/3}$ or if ${\partial_x c_2=-c^2_2/2-2\sigma\Lambda}$. The form of the metric,
\begin{equation}
    \mathrm{d}s^2=\mathrm{d}x^2+2(c_1+c_2r)\,\mathrm{d}u \mathrm{d}x-2\mathrm{d}u \mathrm{d}r+(c_4+c_5r+c_6r^2)\,\mathrm{d}u^2,
\end{equation}
with $c_2$ and $c_6$ given by above equation, is invariant by the coordinate transformations $(r,u,x)\to (r',u',x')$ \cite{Chow:2009vt},
  \begin{equation}\label{eq:ctransf}
   r=\frac{r'}{\dot{u}(u')}+ F(u',x'), \quad u=u(u'),\quad x=x'+G(u'), 
  \end{equation}
together with the following re-definitions
\begin{equation}\label{eq:ctransf3}
    \begin{aligned}
     \dot{u}(c_1+c_2 F)-\dot{u}\frac{\partial F}{\partial x'}+\frac{\mathrm{d}G}{\mathrm{d}u'} &\to c_1,\\
    \dot{u}(c_5+2c_6 F)+2c_2\frac{\mathrm{d}G}{\mathrm{d}u'}+\frac{2\Ddot{u}}{\dot{u}} &\to c_5,\\ 
    \dot{u}^2(c_4+c_5 F+c_6 F^2)-2\dot{u}\frac{\partial F}{\partial u'}+2\dot{u}(c_1+c_2 F)\frac{\mathrm{d}G}{\mathrm{d}u'}+\left(\frac{\mathrm{d}G}{\mathrm{d}u'}\right)^2 &\to c_4,
    \end{aligned}
\end{equation}
where ${\dot{u}=du/du'}$. Since the equation \eqref{eqru} is independent of the coordinate $u$ and translationally symmetric, its solution takes the form $c_2=c_2 (x+c(u))$. Following the coordinate freedom \eqref{eq:ctransf}, the $u$ dependence of the metric function $c_2$ can be eliminated by transformation ${x\to x-c(u)}$. Hence, we can set ${c_2=c_2(x)}$ without loss of generality. The special case ${\partial_x c_2=-c^2_2/2-2\sigma\Lambda}$ then leads to $ c_2=2\sqrt{\sigma\Lambda}\tan [\sqrt{\sigma\Lambda }(2\sigma \tilde{c}-x)]$ where $\tilde{c}$ is a constant. The coordinate freedom \eqref{eq:ctransf} can be used to set ${c_1=0}$ by transforming ${r\to r+F(u,x)}$.

\item \underline{$\mathcal{E}^{\text{TMG}}_{ux}=8\pi T_{ux}$} \newline
With $g_{ux}$ being specified now, we can rewrite energy-momentum conservation equations \eqref{EMTcons} as
\begin{equation}
\begin{aligned}
   & T_{ux}=J,\\&
    T_{uu}=r\left[c_2\,J+\partial_x J\right]+N
    \end{aligned}
    \end{equation}
where the profile functions $J$ and $N$ are again arbitrary functions of $u$ and $x$. The $ux$ component of the field equation can be obtained as
\begin{equation}
 -\frac{1}{\mu} \partial^2_x c_5+\left(\sigma-\frac{c_2}{2\mu}\right)\partial_x c_5=-16\pi  J .
  \label{1.cons}
\end{equation}

\item \underline{$\mathcal{E}^{\text{TMG}}_{uu}=8\pi T_{uu}$} \newline
The $uu$ component of field equations yields
\begin{equation}
\begin{aligned}
    & -\frac{1}{\mu}\partial^3_x c_4+\left(\frac{3}{2\mu}c_2+\sigma\right)\partial^2_x c_4+\left(\frac{3\sigma\Lambda}{\mu} +\frac{14}{4\mu}\partial_x c_2-\sigma c_2+\frac{1}{4\mu}c^2_2\right)\partial_x c_4+\bigg(-\frac{3}{2\mu}c^3_2+ \sigma c^2_2\\
    &-\frac{6\Lambda}{\sigma\mu}c_2-\frac{9}{2\mu}c_2\partial_x c_2
    +4\Lambda+\sigma \partial_x c_2\bigg)c_4-\frac{c_5}{2\mu}\partial_x c_5+\frac{1}{\mu}\partial_x\partial_u c_5=-16\pi  N .
\label{2.cons}
 \end{aligned}
\end{equation}
\end{enumerate}
In conclusion, the Kundt solutions of matter-coupled TMG are given by 
\begin{equation}\label{eq:metricKundtTMG}
    \mathrm{d}s^2=\mathrm{d}x^2+2rc_2 \mathrm{d}u \mathrm{d}x-2\mathrm{d}u\mathrm{d}r+\left[c_4+c_5 r+\frac{1}{4}\left(3c^2_2+4\partial_x c_2+\frac{12\Lambda}{\sigma}\right) r^2\right]\mathrm{d}u^2,
\end{equation}
where $c_2$ is a function of $x$ that is given implicitly as a solution of \eqref{eqru}. The remaining coefficients $c_4$ and $c_5$ are arbitrary functions of $u$ and $x$ that have to satisfy the equations \eqref{1.cons} and \eqref{2.cons}. They can be solved by prescribing the profile functions $J$ and $N$. The function $c_5$ can be found from \eqref{1.cons}, in which it is linear, and substituted in \eqref{2.cons}. This is then a linear equation for $c_4$. Alternatively, if we prescribe metric functions $c_4$ and $c_5$, we can easily determine the profile functions $J$ and $N$. Remark that general vacuum Kundt solutions of TMG were studied in \cite{Chow:2009vt}. Here we generalized these results by including the pure radiation and gyratonic matter. Observe that in the GR limit, the equations \eqref{1.cons} and \eqref{2.cons} reduce to
\begin{equation}
\begin{aligned}
    \sigma\partial_x c_5&=-16\pi J,\\
    -\sigma \partial^2_x c_4+\sigma c_2\partial_x c_4-\left( \frac{\sigma c^2_2}{2}+2\Lambda\right)c_4 &=16\pi  N ,
\end{aligned}
\end{equation}
which matches with the known GR result of \cite{Podolsky:2018zha}.


\subsubsection{Subcase $c_2=0$: Type III pp-wave gyratons}
 
Due to \eqref{eqru}, the case ${c_2=0}$ is compatible only with the vanishing cosmological constant, ${\Lambda=0}$. Assuming so, the metric \eqref{eq:metricKundtTMG} now becomes
\begin{equation}
    \mathrm{d}s^2=\mathrm{d}x^2-2\mathrm{d}u\mathrm{d}r+(c_4+c_5 r)\mathrm{d}u^2.
\end{equation}
and the field equations \eqref{1.cons} and \eqref{2.cons} turn into
\begin{equation}
    \begin{aligned}
      -\frac{1}{\mu} \partial^2_x c_5+\sigma\partial_x c_5 &=-16\pi  J ,
     \\
      -\frac{1}{\mu}\partial^3_x c_4+\sigma\partial^2_x c_4-\frac{c_5}{2\mu}\partial_x c_5+\frac{1}{\mu}\partial_x\partial_u c_5 &=-16\pi  N .
    \end{aligned}
\end{equation}
The general solution in the vacuum ($J=0$ and $N=0$) of the above equations is given by
  \begin{equation}
  \begin{aligned}
        c_5 &=\frac{e^{\sigma\mu x}}{\sigma\mu}d_1(u)+d_2(u)\;,
        \\
        c_4 &=d_3(u)+x d_4(u)-\frac{1}{4\mu^3\sigma}\bigg\{\frac{e^{2\sigma\mu x}}{2\sigma\mu}d^2_1(u)+e^{\sigma\mu x}\big[(2\mu \sigma x-4)(d_1(u)d_2(u)-2d'_1(u))-4\sigma\mu d_5(u)\big]\bigg\}\;.
    \end{aligned}
\end{equation}
The subcase ${c_5=0}$ for the vacuum case was found in \cite{garcía-díaz_2017,Chow:2009km} and a specific massless particle solution was analyzed in \cite{Edelstein:2016nml}.


\subsection{Solving NMG}

Let us now, impose the field equations of NMG.
\begin{enumerate}
    \item \underline{$\mathcal{E}^{\text{NMG}}_{rr}=0$ and $\mathcal{E}^{\text{NMG}}_{rx}=0$ } \newline
The $rr$ component of the field equation yields
\begin{equation}
    \frac{1}{m^2}\left[\partial^4_r g_{uu}+ (\partial^2_r g_{ux})^2-3 \partial_r g_{ux} \partial^3_r g_{ux}-2 \partial_r^3\partial_x g_{ux}-2 g_{ux}\partial^4_r g_{ux}\right]= 0,
    \label{NMG1}
\end{equation}
while the $rx$ component is
\begin{equation}
\begin{aligned}
  & \frac{1}{m^2}\big[ \partial^2_r g_{ux} \left[(\partial_r g_{ux})^2-2 \left(\partial^2_r g_{uu}+4 m^2 \sigma \right)\right]-2 \big[-8\partial_r g_{uu}\partial^3_r g_{ux}+\partial_r g_{ux} \left(5\partial^3_r g_{uu}+6 g_{ux}\partial^3_r g_{ux}\right)
  \\
  &+2\partial_r\partial_x g_{uu}+4 g_{ux}\partial^3_r\partial_x g_{ux}-8 \partial^3_r\partial_u g_{ux}\big]+8
   \partial^4_r g_{ux} \left(g_{uu}-g_{ux}^2 \right)+4 g_{ux}
   (\partial_r g_{ux})^2\big]=0.
   \end{aligned}
   \label{NMG2}
\end{equation}
Again, in order to solve \eqref{NMG1} and \eqref{NMG2}, we will additionally assume that $g_{ux}$ is a polynomial function in $r$. As discussed in \cite{Chakhad:2009em}, $g_{ux}$ then actually needs to be linear in $r$,
\begin{equation}
    g_{ux}= c_1+c_2 r,
\end{equation}
where $c_1$ and $c_2$ are arbitrary functions of coordinates $u$ and $x$. Consequently, the $rr$ component of field equation \eqref{NMG1} turns into
\begin{equation}
    \partial^4_r g_{uu}= 0,
\end{equation}
solution of which is
\begin{equation}
    g_{uu}= c_3+c_4 r+c_5 r^2+c_6 r^3,
\end{equation}
where $c_3$, $c_4$, $c_5$, and $c_6$ are other arbitrary functions of coordinates $u$ and $x$. The $rx$ component of the field equation \eqref{NMG2} simplifies to
\begin{equation}
    2\partial_x c_6+5c_2 c_6=0.
\end{equation}
Since the second derivative of the trace of the field equations with respect to $r$ implies ${c_6=0}$, the previous equation is actually trivially satisfied. Also, as in TMG case, we can again set ${c_1=0}$ [by means of the freedom ${r\to r+F(u,x)}$].

\item \underline{$\mathcal{E}^{\text{NMG}}_{xx}=0$}  \newline
The $xx$  component of the field equation is
\begin{equation}
  -\frac{1}{m^2}\left[c^4_2+c^2_2(-16\sigma m^2+88\,c_5)+160c_2\partial_x c_5+16(-4m^2\Lambda_0+c^2_5+4c_5(\sigma m^2+\partial_x c_2)+4\partial^2_x c_5)\right]=0.
  \label{NMGxxcomp}
\end{equation}
\item \underline{$\mathcal{E}^{\text{NMG}}_{ru}=0$}  \newline
The $ru$  component of the field equation can be given
\begin{equation}
 \frac{1}{m^2}\left[ c^4_2+4c^2_2(-4\sigma m^2-12\,c_5+\partial_x c_2)-80c_2\partial_x c_5-16(c^2_5+3c_5\partial_x c_2+4m^2\Lambda_0+2\sigma m^2\partial_x c_2+2\partial^2_x c_5)\right]=0.
  \label{eqruNMG}
\end{equation}
On the other side, the trace of the field equations leads to
\begin{equation}
  \frac{1}{m^2}\left[-3c^4_2+8c^2_2(6\sigma m^2+c_5-\partial_x c_2)+16(c^2_5+2c_5(-2m^2\sigma+\partial_x c_2)+4m^2(3\Lambda_0+\sigma \partial_x c_2))\right]=0 
  \label{eqTrNMG}
\end{equation}
which can be also obtained as $\mathcal{E}^{\text{NMG}}_{xx}-2\mathcal{E}^{\text{NMG}}_{ru}$. Observe that, the two equations \eqref{eqruNMG} and \eqref{eqTrNMG} are independent of the coordinate $u$ and translationally symmetric, hence the solution must take the form ${c_2=c_2 (x+c(u))}$ and ${c_5=c_5 (x+c(u))}$.  Therefore, $u$ dependence of the metric functions $c_2$ and $c_5$ can be removed by the coordinate transformation $x\to x-c(u)$; namely we may set ${c_2=c_2(x)}$ and ${c_5=c_5(x)}$. 

\item \underline{$\mathcal{E}^{\text{NMG}}_{ux}=8\pi T_{ux}$} \newline
Again, the energy-momentum conservation equations \eqref{EMTcons} now allow us to write
\begin{equation}
\begin{aligned}
   & T_{ux}=J,\\&
    T_{uu}=r[c_2\,J+\partial_x J]+N,
    \end{aligned}
    \end{equation}
where $J$ and $N$ are arbitrary functions of $u$ and $x$. The $ux$ component of the field equation yields
\begin{equation}
  \frac{1}{m^2}\left[8\,\partial^3_x c_4+8\,c_2\,\partial^2_x c_4+(c^2_2+4c_5+4\partial_x c_2)\partial_x c_4\right]+8\sigma \partial_x c_4 =-128\pi J.
  \label{1.consNMG}
\end{equation}

\item \underline{$\mathcal{E}^{\text{NMG}}_{uu}=8\pi T_{uu}$} \newline
The $uu$ component of the field equation is
\begin{equation}
\begin{aligned}
& \frac{1}{m^2}\Big\{\left[16c^2_5-\partial_x c_2(8 m^2\sigma+11 c_2^2)+36(\partial_x c_2)^2+4c_5(8m^2\sigma+17c^2_2+11\partial_xc_2)+36c_2(3\partial_xc_5+\partial^2_xc_2)-24\partial^2_xc_2\right]c_3\\&+\left[-3c^3_2+6c_2(-4m^2\sigma-2c_5+15\partial_xc_2)-84\partial_xc_5-84\partial^2_xc_2\right]\partial_xc_3+\left[27c^2_2+12(2m^2\sigma-3c_5-9\partial_xc_2)\right]\partial^2_xc_3\\&
+24c_4\partial^2_xc_4+24(\partial_xc_4)^2-48c_2\partial^3_xc_3+24\partial^4_xc_3-48\partial_u\partial^2_xc_4\Big\}=-384\pi N.
\label{2.consNMG}
 \end{aligned}
\end{equation}
 \end{enumerate}
In conclusion, Kundt solution of matter-coupled NMG reads 
\begin{equation}
    \mathrm{d}s^2=\mathrm{d}x^2+2rc_2 \mathrm{d}u \mathrm{d}x-2\mathrm{d}u\mathrm{d}r+\left(c_3+c_4 r+c_5 r^2\right)\mathrm{d}u^2 ,
\end{equation}
where $c_3$ and $c_4$ are functions of $u$ and $x$,  while $c_2$ and $c_5$ are functions of $x$. These functions must satisfy the set of equations \eqref{NMGxxcomp}, \eqref{eqruNMG}, \eqref{1.consNMG} and \eqref{2.consNMG}.  If we prescribe $c_2$, $c_3$ and $J$, we can integrate the equation \eqref{NMGxxcomp} and \eqref{1.consNMG} to obtain $c_5$ and $c_4$ respectively. Then,  \eqref{2.consNMG} gives profile function $N$. Kundt solutions of source-free NMG found in \cite{Chakhad:2009em}. Above results generalize the solutions to the presence of pure radiation and gyratonic matter. Note that in the GR limit,  \eqref{NMGxxcomp} and \eqref{eqruNMG} are trivially satisfied, and the equation \eqref{1.consNMG} and \eqref{2.consNMG} turn into
\begin{equation}
\begin{aligned}
      \sigma \partial_x c_4 &=-16\pi J,\\
     -\sigma \partial^2_x c_3+\sigma c_2\partial_x c_3-\left( \frac{\sigma c^2_2}{2}+2\Lambda\right)c_3 &=16\pi  N ,
\end{aligned}
\end{equation}
which are the correct 3D GR result \cite{Podolsky:2018zha}.


\subsubsection{Subcase $c_2=0$: Type III pp-wave gyratons in NMG}
In this case, we have $\Lambda_0=-m^2$ (the bound \eqref{eq:Lambdapm} is saturated ) which leads to $c_5=-2\sigma m^2$, the metric takes the following form
\begin{equation}
    \mathrm{d}s^2=\mathrm{d}x^2-2\mathrm{d}u\mathrm{d}r+\left(c_3+c_4 r-2\sigma m^2 r^2\right)\mathrm{d}u^2,
\end{equation}
and the functions are constrained by the following equations 
\begin{equation}
    \begin{aligned}
      \frac{1}{m^2} \partial^3_x c_4&=-16\pi  J,
     \\
      4m^2\sigma\partial^2_xc_3
+c_4\partial^2_xc_4+(\partial_xc_4)^2 
+\partial^4_xc_3-2\partial_u\partial^2_xc_4&=-16\pi m^2 N.
\label{type3pp}
    \end{aligned}
\end{equation}
These equations provide a connection between profile functions and metric functions $c_3$ and $c_4$. Unlike the 3D GR and TMG cases, the equations depend on the cosmological constant. Some explicit solutions can be constructed using \eqref{type3pp}. For the sake of simplicity, let us consider the vacuum case. In this case, the solution can be found as
\begin{equation}
 \begin{aligned}
    \mathrm{d}s^2&=\mathrm{d}x^2-2\mathrm{d}u\mathrm{d}r+\Big\{\bigg[-\frac{d_3^2}{8\sigma m^2}x^4-\frac{d_2 d_3}{4\sigma m^2} x^3+(\frac{3 d_3^2}{8m^4}-\frac{ d_2^2+2d_1 d_3-4\partial_u d_3}{8\sigma m^2}) x^2+d_4+x d_5\\&-\frac{1}{4\sigma m^2}\left(d_6\cos(2\sqrt{\sigma}m x)+d_7\sin(2\sqrt{\sigma}m x)\right)\bigg]+(d_1+d_2x+d_3x^2) r-2\sigma m^2 r^2\Big\}\mathrm{d}u^2,
      \end{aligned}
\end{equation}
where $d_i$'s are functions of the coordinate $u$. Note that this solution is not a solution for 3D GR.


\section{Conclusions}\label{sc:conclusions}

In the present work, we analyzed exact solution of TMG and NMG in the presence of cosmological constant that belongs to the (expanding) Robinson-Trautman metrics with the GR-like expansion ${\Theta=1/r}$ and the Kundt metrics ($\Theta=0$). We considered the energy-momentum tensor with pure radiation component $T_{uu}$ and gyratonic matter component $T_{ux}$. We found the general solutions of TMG. For NMG, we constructed solutions with an additional assumption of $g_{ux}$ being polynomial in $r$ (this condition is automatically enforced in 3D GR and TMG). 

We showed that the metric functions for GR-like Robinson-Trautman are constrained by three field equations for TMG and four field equations for NMG (in contrast to two field equations in the case of 3d GR). The metric function $c_1$ remains arbitrary in both cases (similar to 3D GR) whereas it is zero in dimensions higher than three. In this sense, gravity is ´less constrained' in 3D because it admits more solutions. On the other hand, in the case of ${c_1=0}$ and ${P=1}$, we found the gravitational field of the Vaidya-type object which emits pure radiation. In each case, we also gave an explicit solution which is not a solution of 3D GR.  

In the Kundt class ($\Theta=0$), the off-diagonal metric function $c_2$ is not zero in general. We showed that solutions of 3D GR are not solutions to these theories. We also identified type-III pp-wave gyratons and studied solutions for the specific source terms. 

Similar analysis can be done for other 3D modified gravity theories such as (generalied) MMG, etc.   


\section{\label{ackno} Acknowledgements}

The authors would like to thank Ji\v{r}\'i Podolsk\'y, Bayram Tekin, Robert \v{S}varc, and Tom\'a\v{s} M\'alek for stimulating discussions. I.K. was supported by Primus grant PRIMUS/23/SCI/005 from Charles University. E.K. was partially supported by the TUBITAK Grant No. 119F241.


%


\end{document}